Communication / Letter

# NMR surface relaxivity in a time-dependent porous system

Neil Robinson, Razyq Nasharuddin, Einar O. Fridjonsson and Michael L. Johns*

*Department of Chemical Engineering, The University of Western Australia, 35 Stirling Highway, Perth WA 6009, Australia*

***Corresponding Author:**
Professor Michael L. Johns
Chair of Chemical and Process Engineering
The University of Western Australia

**Postal Address:**
Department of Chemical Engineering
The University of Western Australia
35 Stirling Highway (M050)
Perth WA 6009
Australia

**Phone:** +61 (08) 6488 5664
**Email:** michael.johns@uwa.edu.au

**Fluid Science and Resources Research Group:** www.fsr.ecm.uwa.edu.au

**ORCID:**

| | |
|---|---|
| Neil Robinson | 0000-0002-0893-2190 |
| Einar O. Fridjonsson | 0000-0001-8365-6002 |
| Michael L. Johns | 0000-0001-7953-0597 |

## Abstract

We demonstrate an unexpected decay-recovery behaviour in the time-dependent $^1$H NMR relaxation times of water confined within a hydrating porous material. Our observations are rationalised by considering the combined effects of decreasing material pore size and evolving interfacial chemistry, which facilitate a transition between surface-limited and diffusion-limited relaxation regimes. Such behaviour necessitates the realisation of temporally evolving surface relaxivity, highlighting potential caveats in the classical interpretation of NMR relaxation data obtained from complex porous systems.

## Main Text

While functional porous materials underpin a vast array of processes of importance to the energy, environment, chemical and construction sectors, characterisation of key structural and interfacial properties within such systems is severely hampered by their optically opaque nature. Nuclear magnetic resonance (NMR) relaxation measurements (also termed nuclear spin relaxation measurements) provide a versatile and non-destructive approach with which to probe the dynamics of spin-bearing fluids within porous materials [1,2], and have been applied widely to the study of both equilibrium fluid properties (informing pore size distributions [3,4], adsorption phenomena [5,6] and diffusive exchange processes [7,8]) and to obtain time-resolved insight into evolving material structures (such as cements [9–13]). In this letter we expand upon the established interpretation of NMR relaxation data when probing such time-resolved material properties, elucidating the uniquely coupled sensitivity of such measurements to the temporal evolution of both pore structure characteristics and surface chemistry properties simultaneously.

For fluid-saturated porous media, expressions for the dependence of observed $^1$H (proton) longitudinal ($T_1$) and transverse ($T_2$) NMR relaxation behaviour on pore structure and interfacial chemistry are well-known, taking the general form [14]:

$$\frac{1}{T_i} \approx \frac{1}{T_{i,bulk}} + \left(\frac{d}{2\alpha\rho_i} + \frac{d^2}{8\alpha D}\right)^{-1}, \quad (1)$$

wherein $i \in \{1,2\}$, and where additional terms may be required to fully account for observed $T_2$ relaxation rates in the presence of magnetic susceptibly contrast across the solid/fluid interface [15,16]; such effects are mitigated in this work by performing our measurements at low magnetic field [17]. Here, $T_i$ are the observed (measured) time constants and $T_{i,bulk}$ represent the time constants for the unrestricted bulk fluid. Terms within parentheses then describe the extent to which pore structure and interfacial chemistry perturb the observed relaxation characteristics: $d$ is the pore diameter, $\alpha$ is a dimensionless shape parameter (taking values of 1, 2 or 3 for planar, slit or cylindrical pores, respectively), and $D$ is the self-diffusion coefficient of the confined fluid. The terms $\rho_i$ are the (spatially averaged) surface relaxivities of the solid/fluid interface, which may be modelled as $\rho_i = \lambda/T_{i,surf}$ [18], where $T_{i,surf}$ are the relaxation time constants within an adsorbed surface layer of thickness $\lambda$. Such terms describe enhanced rates of relaxation which occur at solid/fluid interfaces both due to a reduction in molecular mobility within the adsorbed surface layer [19] and through dipolar proton-electron interactions between adsorbate-bound $^1$H and paramagnetic species on the pore surface [14]. Established limiting cases for such relaxation dynamics exist. In the limit $D/d \ll \rho_i$ for instance, surface relaxation rates are significantly more rapid than the rates of diffusive transport across the pores, with **Equation (1)** reducing to:

$$\frac{1}{T_i} \approx \frac{1}{T_{i,bulk}} + \frac{8\alpha D}{d^2}. \quad (2)$$

Conversely, in the limit $D/d \gg \rho_i$, diffusion across the pore is sufficiently more rapid than the rates of enhanced surface relaxation at the pore surface, with **Equation (1)** reducing to:

$$\frac{1}{T_i} \approx \frac{1}{T_{i,bulk}} + \frac{2\alpha\rho_i}{d}. \quad (3)$$

Such limits are referred to according to the overall rate controlling process; **Equation (2)** therefore describes diffusion-limited relaxation, while **Equation (3)** describes surface-limited relaxation. Here, we discuss data which for the first time permits clear identification of a temporal transition between these limiting regimes within an evolving three-dimensional porous micro-structure.

**Figure 1** shows $^1$H $T_1$ and $T_2$ distributions obtained at low magnetic field ($B_0$ = 0.05 T; $\nu_0$($^1$H) = 2 MHz) for tap water within the hydrating engineering material cemented paste backfill (CPB) [20–22]. This porous material comprised a mixture of spherical fly ash particles, minerals tailings, Ordinary Portland cement and water (see Supplementary Information for extended details of materials preparation), with the solids containing 7.9 wt%, 3.2 wt%, and 2.6 wt% $Fe_2O_3$, respectively, as measured by inductively coupled plasma optical emission spectroscopy (ICP-OES), wherein paramagnetic $Fe^{3+}$ ions provide the dominant source of relaxation sinks for adsorbed water at the pore surface. NMR measurements were performed under ambient conditions using inversion recovery [23] and CPMG (Carr-Purcell Meiboom-Gill) NMR pulse sequences [24,25], respectively, with the resulting data inverted to produce probability density distributions of observed $T_1$ and $T_2$ times via Tikhonov regularisation [26–28] (see Supplementary Information for extended NMR methods, including Refs. [29–33]). Acquired $T_1$ distributions (**Figure 1a**) show a large primary peak, typical of systems in which sufficiently slow $T_1$ relaxation characteristics allow extensive diffusive mixing between water confined within different pore sizes [34]. Acquired $T_2$ distributions (**Figure 2a**), which are naturally more sensitive to local pore geometry than $T_1$ due to shorter relaxation times, reveal multiple relaxation environments, indicative of water confined within a hierarchically porous cement structure. Such distributions are consistent with established models of cement hydration [35], with the assignment of these relaxation peaks to specific hydrating pore structures detailed elsewhere [36,37].

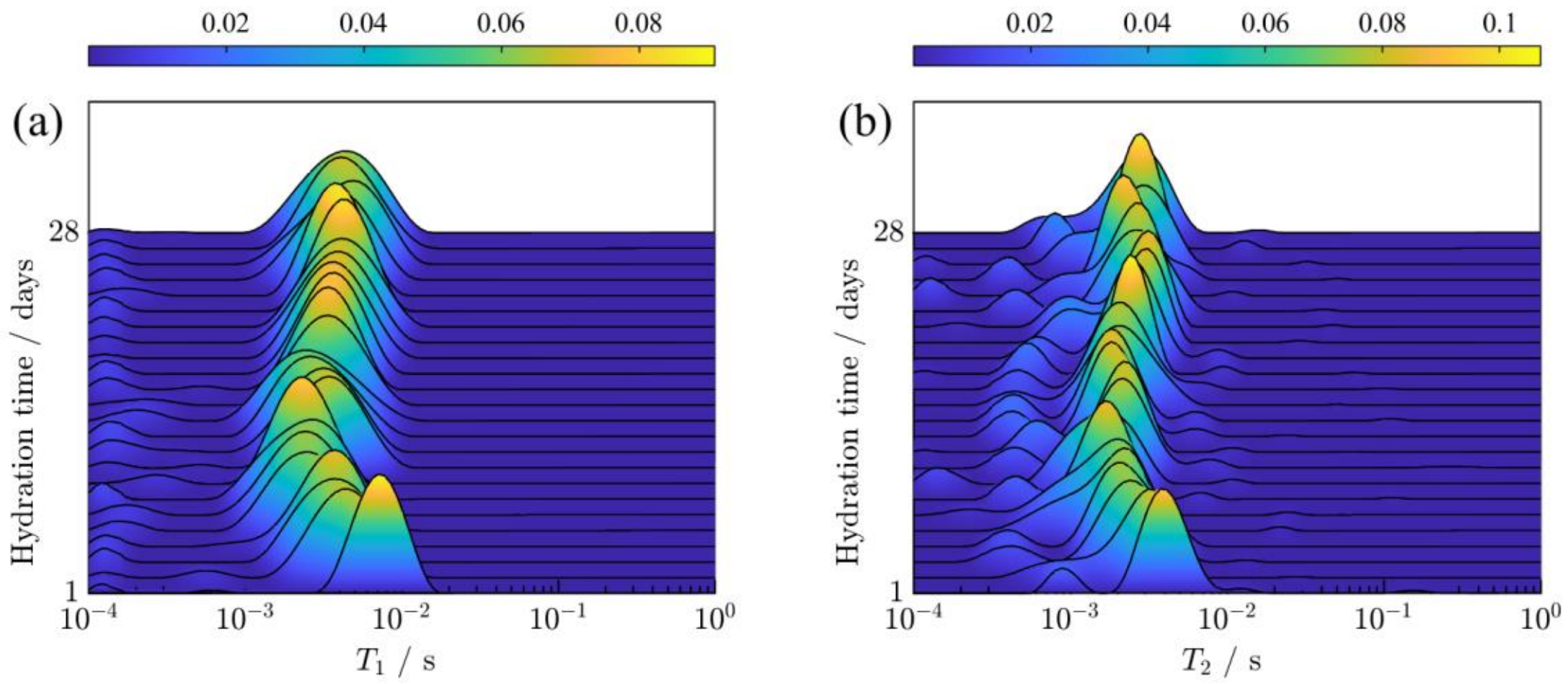

**Figure 1.** Example (a) $T_1$ and (b) $T_2$ relaxation distributions for the hydrating material investigated in this study. Relaxation data was acquired over the first 28 days of material hydration. Colour bars define the probability density of each relaxation distribution.

In this letter we consider only the modal relaxation times from each inverted distribution, termed $\langle T_1 \rangle$ and $\langle T_2 \rangle$, respectively, which are dominated by the most populous pore structures within the hydrating material under study (capillary pores in traditional cement chemistry notation [38]). In recent work investigating the hydration behaviour of a similar material in the absence of fly ash particles, a clear mono-exponential relationship was observed between these modal time constants and hydration time $t_h$, which defines the experimental period across which the material is allowed to evolve. This relationship took the form $\langle T_i \rangle_{sl} = a_i \exp(-b_i t_h) + c_i$ [36], where $b_i$ are the observed material hydration rates, $a_i$ provides a scaling factor, and $c_i$ are offset parameters necessary to account for non-zero pore sizes at long $t_h$; the subscript '$sl$' indicates that surface-limited relaxation is assumed within this system [37]. Fitting $\langle T_i \rangle_{sl}$ against $t_h$ then enables the extraction and comparison of material hydration rates $b_i$, facilitating quantitative contrast between different material formulations, preparation procedures, and hydration conditions.

**Figures 2a and 2b** detail the temporal evolution of our acquired $\langle T_1 \rangle$ and $\langle T_2 \rangle$ data across 28 days of hydration. Rather than the expected mono-exponential decay behaviour, however, a clear decay-recovery process (subscript '$dr$') is observed for each data set which may be modelled as:

$$\langle T_i \rangle_{dr} = a_i' \exp(-b_i' t_h) - a_i \exp(-b_i t_h) + c_i. \qquad (4)$$

We rationalise this bi-exponential behaviour through consideration of a transition between the diffusion- and surface-limited relaxation processes described in **Equations (2) and (3)**. In our chosen material the use of fly ash introduces a high $Fe_2O_3$ content (7.9 wt% as measured by ICP-OES), providing a significantly increased paramagnetic $Fe^{3+}$ concentration ([$Fe^{3+}$]) compared to that present in previous relaxation investigations of similar hydrating materials [36,37]. Surface relaxivity values $\rho_i = \lambda/T_{i,surf}$ are sensitive to [$Fe^{3+}$] via the established expressions [10,14]:

$$\frac{1}{T_{1,surf}} = \beta[3J(\omega_I) + 7(\omega_S)]$$
$$\frac{1}{T_{2,surf}} = \frac{\beta}{2}[8J(0) + 3J(\omega_I) + 13(\omega_S)], \quad (5)$$

where the spectral density functions $J(\omega_x) = \tau_m \ln[1 + \omega_x^2\tau_m^2/(\tau_m^2/\tau_s^2) + \omega_x^2\tau_m^2]$ (with $x \in \{I, S\}$) capture the relationship between surface $^1H$ relaxation rates and the dynamics of adsorbed water molecules at the internal pore surfaces of the hydrating material [14]. Here, the correlation times $\tau_m$ and $\tau_s$ are the translational correlation time of water molecules between paramagnetic relaxation sinks, and the surface residence time before desorption, respectively, while the frequencies $\omega_I$ and $\omega_S$ are obtained via $\omega_x = \gamma_x B_0$, where $\gamma_I$ and $\gamma_S$ are the proton and electron gyromagnetic ratios, respectively [39]. Importantly, the prefactor is given by $\beta \propto \sigma\delta^{-4}\gamma_I^2\gamma_S^2\hbar^2 S(S+1)$, where $\sigma$ is the density of paramagnetic surface species of spin $S$ at the pore surface (and hence assumed directly dependent on [$Fe^{3+}$] for our system; $S = 5/2$ for $Fe^{3+}$); $\delta$ is the distance of closest approach between adsorbate molecule and surface spins, while $\hbar$ is the reduced Plank constant [10].

Godefroy *et al.* demonstrated a dependence between [$Fe^{3+}$] and limiting relaxation regime for a series of model porous structures comprising silicon carbide grains [14,40], evidencing a clear transition between diffusion-limited and surface-limited relaxation upon material cleaning to reduce [$Fe^{3+}$]. In the present work we conjecture than the temporal evolution of our hydrating porous material serves to facilitate this transition. Specifically, the dissolution-precipitation reactions responsible for cement hydration processes lead to the deposition of calcium silicate hydrate (C-S-H in cement chemistry notation) adlayers at the material pore surface [41], which is expected to both decrease the capillary pore size and limit access to the highly paramagnetic fly ash particles upon increasing hydration time. We support this interpretation in **Figure 2** via scanning electron microscope (SEM) imaging of a sample of our hydrating material, which clearly demonstrates the time-dependent deposition of C-S-H across fly ash particle surfaces during the hydration period investigated. At short $t_h$ we therefore interpret our observed relaxation behaviour as dominated by interactions with high [$Fe^{3+}$] fly ash particles (**Figure 2c**); following **Equation (5)** these interactions facilitate large $\rho_i$ values, causing pore water to undergo diffusion-limited $^1H$ relaxation (subscript '$dl$') according to **Equation (2)**. A result of the above interpretation is that a square dependence on pore size is expected during pore structure evolution at short $t_h$, of the form $T_i \propto d^2$. Given the established mono-exponential decay relationship between $\langle T_i \rangle_{sl}$ and $t_h$ in the absence of fly ash co-binder (of the form $\langle T_i \rangle_{sl} \sim d \sim \exp(-b_i t_h)$ [36]), an initial decay of the form $\langle T_i \rangle_{dl} \sim d^2 \sim \exp(-2b_i t_h)$ is then expected from our present data; this rapid decay processes is accounted for by the primed components of our bi-exponential model expression in **Equation (4)**. As a simple visual comparison, blue dotted decay curves within **Figures 2a and 2b** show a scaled mono-exponential fit to our $\langle T_1 \rangle$ and $\langle T_2 \rangle$ data at $t_h$ < 10 days (with decay rates constrained to equal those obtained from our full fit to **Equation (4)** described below), while fitted relaxation data obtained from a comparable hydrating material in the absence of highly paramagnetic fly ash are shown in purple (data obtained from reference [36]; further details are provided in the Supplementary Information). Enhanced relaxation time decay rates are clearly evident in the presence of fly ash (blue dotted curves), providing support for the expectation of comparably large $\rho_i$ values, and hence for the interpretation of diffusion-limited relaxation at short $t_h$. Note that given these two hydrating materials are likely characterised by different pore structures and surface relaxivities, quantitative comparisons between the relaxation time values exhibited by these two systems are not possible, and we make no attempt to ascribe a physical interpretation to the convergence of these curves at short $t_h$, which can only be coincidental.

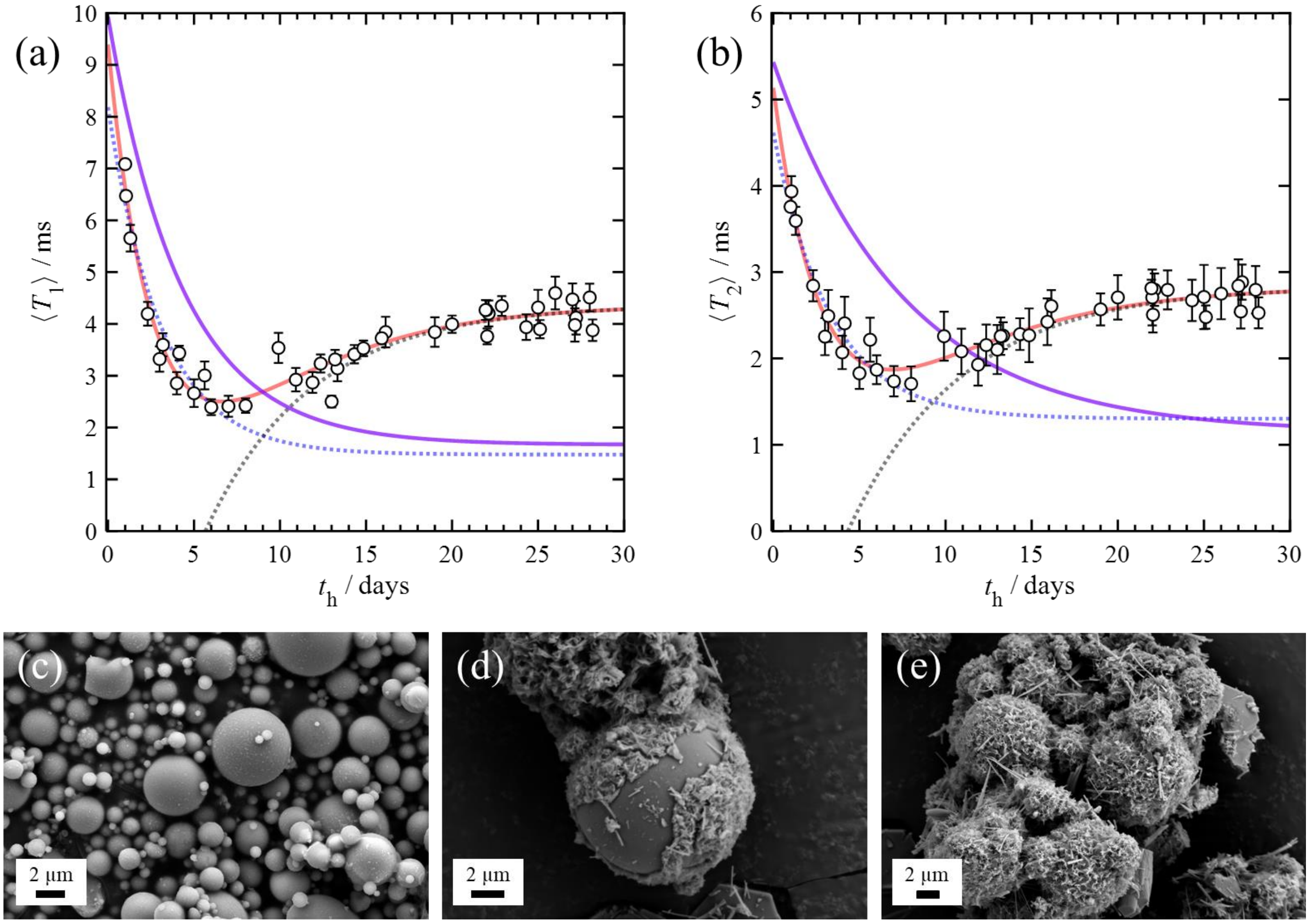

**Figure 2.** (a) $\langle T_1 \rangle$ and (b) $\langle T_2 \rangle$ $^1$H NMR relaxation evolution of water confined within the hydrating CPB material investigated here over 28 days. Error bars indicate ± 1 standard deviation (n = 3), while data points without error bars were acquired only once. Red curves indicate a fit to **Equation (4)** with $b_i' = 2b_i$. Blue curves show mono-exponential fits to data at $t_h$ <10 days, while grey curves represent the deconvoluted recovery components of **Equation (4)**, respectively. Purple curves show fitted mono-exponential decay data obtained from similar hydrating systems in the absence of highly paramagnetic fly ash [36], demonstrating reduced decay rates relative to our data at short $t_h$. Bottom panels illustrate SEM images of fly ash particles; (c) shows fly ash particles in the absence of any cement hydration products ($t_h$ = 0), while (d) and (e) evidence surface C-S-H adlayer formation following 7 and 14 days of hydration, respectively.

Following nucleation at the solid/liquid interface, C-S-H deposits continue to grow across the available pore surfaces according to established mechanisms of cementitious hydration [35]. The evolution of such adlayers is demonstrated for the present system in **Figures 2d and 2e**, which reveal the difference in adlayer growth following $t_h = 7$ days and $t_h$ = 14 days, respectively. As C-S-H adlayers cover the internal surfaces of the hydrating pore network, dipolar interactions between the high concentration of fly ash-bound $Fe^{3+}$ and $^1H$ spins associated with confined pore water will be reduced, decreasing the spatially averaged surface relaxivity $\rho_i$ of the capillary pore structures; such continued adlayer growth is therefore fundamental in initiating the proposed transition between diffusion- and surface-limited relaxation. An associated transition time may be estimated and will be associated with the shortest measured $\langle T_1 \rangle$ and $\langle T_2 \rangle$ times, which in the present work occur at $t_h \approx 7$ days. While the continuous temporal growth and deposition of surface adlayers serves to reduce the material pore sizes over time, this process continues to be convoluted with the above reduction in $\rho_i$. Such pore size changes do not lead to a continuous reduction in $\langle T_i \rangle_{dr}$ as would occur within systems exhibiting classical surface-limited relaxation wherein $\rho_i$ values are assumed time-independent ($d\rho_i/dt_h \approx 0$). Rather, we interpret the observed *increase* in $\langle T_i \rangle_{dr}$ with increasing $t_h$ as a direct consequence of the coupling between pore structure evolution and decreasing $\rho_i$; this $\langle T_i \rangle$ recovery process is deconvoluted in grey within **Figures 2a and 2b** and is accounted for by the second exponential term of **Equation (4)**.

A consequence of the above discussion is that we may now directly interpret $b_i$ within **Equation (4)** as material hydration rates, which are equated with the rate of changing surface relaxivity under surface-limited conditions. Appropriate fitting of our data to this expression with the constraint $b_i' = 2b_i$ was performed (red curves in **Figures 2a and 2b**), yielding hydration rates $b_1 = 0.16 \pm 0.02$ day$^{-1}$ and $b_2 = 0.15 \pm 0.02$ day$^{-1}$. Remarkably, these values are in excellent agreement with previously observed CPB hydration rates of 0.08 – 0.2 day$^{-1}$ obtained in the absence of fly ash [36]. This result provides both confidence in the above data interpretation and suggests that the assumption of $d\rho_i/dt_h \approx 0$ in the absence of fly ash is reasonable. We note for completeness that while $b_1 \approx b_2$ here, such equivalence is not necessarily expected given the differing sensitivities of $\langle T_1 \rangle$ and $\langle T_2 \rangle$ to the fine structure of hierarchical pore networks within hydrating cement-like materials (demonstrated in **Figure 1**).

In summary, this work has demonstrated the measurement of nuclear spin relaxation phenomena within a dynamic and temporally evolving three-dimensional porous structure exhibiting changes to both pore size and pore surface chemistry. Our observations provide clear evidence that the $^1H$ NMR relaxation characteristics of water confined within such systems can exhibit sensitivity to both phenomena simultaneously, and that our measured relaxation data is susceptible to the resulting temporal changes in pore structure surface relaxivity. We anticipate the consideration of such time dependent surface relaxation phenomena will be critical in avoiding the erroneous interpretation of time-resolved relaxation data obtain from complex porous materials, especially when considering systems containing potentially high concentrations of paramagnetic species, or in reactive (catalytic) systems, where e.g. the coking of polar pore surfaces is expected through the formation of reaction by-products.

## Acknowledgements

The authors thank Prof Andy Fourie (The University of Western Australia) for providing materials. R.N. acknowledges support from an Australian Government Research Training Scholarship and N.R. acknowledges support from the Forrest Research Foundation. The authors further acknowledge use of the Australian Microscopy & Microanalysis Research Facility at The University of Western Australia Centre for Microscopy, Characterisation and Analysis.

# Supplementary Information

# Evolving NMR surface relaxivity in a time-dependent porous system

Neil Robinson, Razyq Nasharuddin, Einar O. Fridjonsson and Michael L. Johns*

*Department of Chemical Engineering, The University of Western Australia, 35 Stirling Highway, Perth WA 6009, Australia*

*michael.johns@uwa.edu.au

## 1) Supplementary Materials and Methods

### Cemented paste backfill preparation

The hydrating porous material employed in this work was cemented paste backfill (CPB), which is an environmentally favourable and long term solution for the deposition of mine tailings (waste solid materials which arise following the processing of mine ores). CPB comprises a mixture of tailings with hydraulic binders such as cement, slag and/or fly ash.

The tailings materials utilised in this investigation were collected from a gold mine in Western Australia. The tailings were sieved (< 4.75 mm) to separate any coagulated particles formed during storage and dehydrated by oven drying at 110 °C for 24 hours before use. Figure S1 shows the particle size distribution of both the tailings and commercially available fly ash (sourced from BGC Cement, Australia) used in this study, which exhibited mean ($d_{50}$) particle sizes of 60 μm and 9 μm, respectively. Mineralogical compositions are outlined in Table S1 and were attained via Inductively Coupled Plasma-Optimal Emission Spectroscopy (ICP-OES). Experimental procedures for our ICP-OES measurements are provided in detail elsewhere [1, 2].

A mixture of fly ash and ordinary Portland cement (OPC) (also sourced from BGC Cement, Australia, **see Table S1** for composition) in a 2:1 ratio by mass made up the binder of our CPB; the practical application of the inclusion of fly ash is to lower economic costs. We opted for a binder content of 9 wt% relative to the total tailings and binder mass, since binder contents between 1 and 10 wt% are common in mine sites [3]. As in our previous CPB-based publications, we maintained a total solids content of 76 wt% for all CPB samples investigated [1]. The water/binder ratio was 3.5. To produce each sample, dehydrated tailings and hydraulic binders were mixed with common tap water for 3 minutes at 90 rpm using a standard stand mixer. The resultant CPB pastes were then transferred to a cylindrical glass vial (diameter = 25 mm, height = 60 mm) for nuclear magnetic resonance (NMR) analysis. All samples were stored at 20 ± 2 °C with a relative humidity of 95 % until the desired hydration time was reached, as per previously published CPB preparation procedures [4].

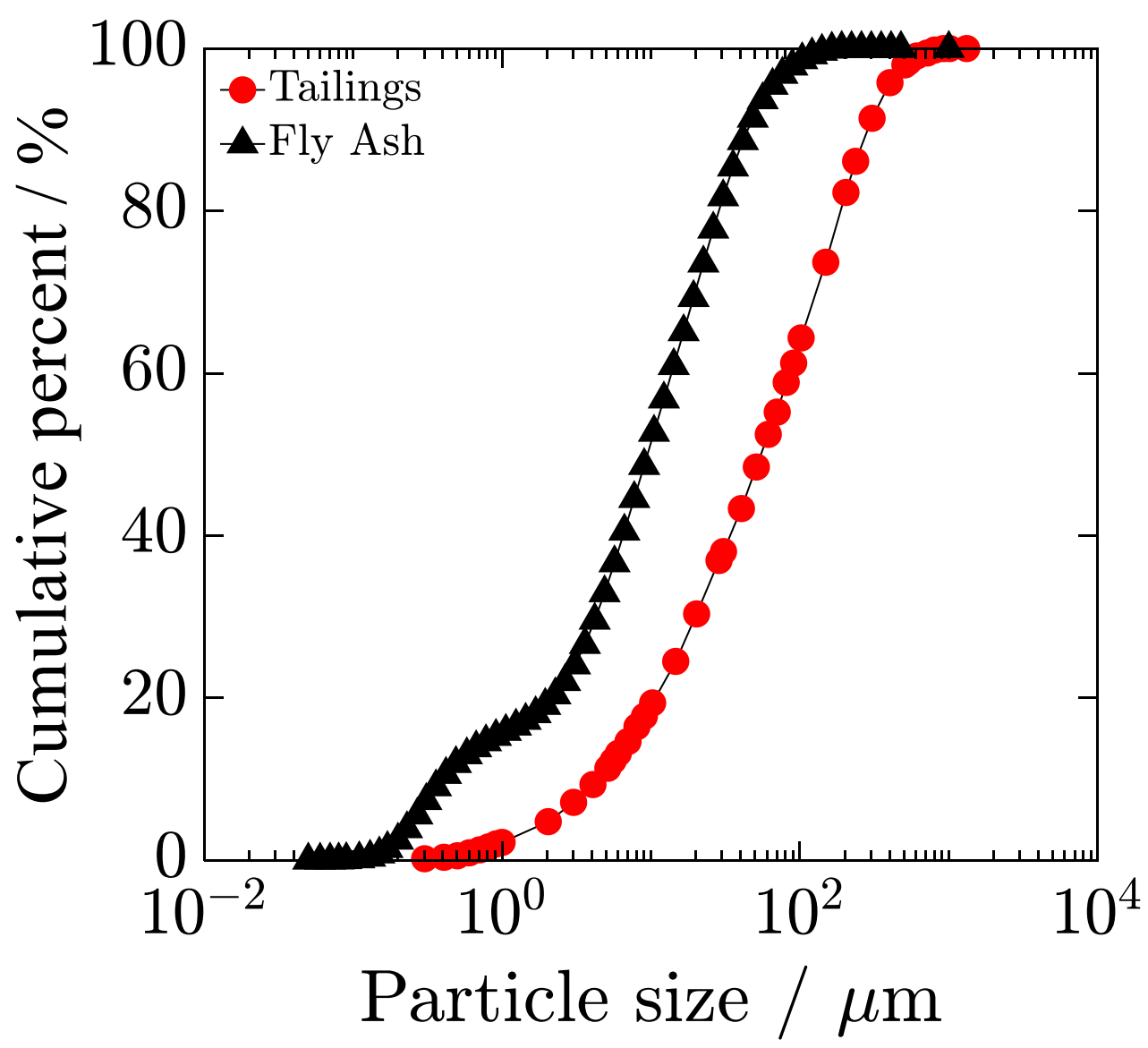

**Figure S1**: Particle size distributions of tailings and fly ash. Tailings exhibit $d_{60}$ ~ 90 µm, $d_{50}$ ~ 60 µm and $d_{10}$ ~ 5 µm, while the fly ash exhibits $d_{60}$ ~ 14 µm, $d_{50}$ ~ 9 µm, and $d_{10}$ ~ 0.4 µm. Prior to analysis both the tailings and fly ash were pre-screened to remove large rocks and agglomerates.

**Table S1**: Chemical composition (obtained from ICP-OES) of the mine tailings, fly ash and OPC (both commercially sourced from BGC Cement, Australia) used in this study.

| **Chemical composition** | **Tailings (wt%)** | **Fly ash** | **Ordinary Portland Cement (OPC)** |
|---|---|---|---|
| $SiO_2$ | 68.49 | 55.90 | 21.00 |
| $Al_2O_3$ | 9.55 | 23.93 | 5.26 |
| $Na_2O$ | 3.31 | 0.38 | 0.35 |
| $Fe_2O_3$ | 3.18 | 7.94 | 2.57 |
| CaO | 3.16 | 6.98 | 65.11 |
| MgO | 1.56 | 1.27 | 1.39 |
| $K_2O$ | 1.34 | 1.02 | 0.43 |
| $SO_3$ | 1.13 | 0.30 | 3.03 |
| $TiO_2$ | 0.54 | 1.31 | 0.26 |
| MnO | 0.08 | 0.10 | 0.08 |
| $P_2O_5$ | 0.05 | 0.54 | 0.10 |

## NMR measurements

Nuclear magnetic resonacne (NMR) relaxation time measurements were conducted using a Magritek Rock Core Analyser equipped with a 0.05 T cylindrical Halbach magnet array, providing a $^{1}H$ NMR Larmor frequency of 2 MHz. Radio frequency (RF) pulse lengths were fixed at 20 μs, with pulse amplitudes for 90° and 180° RF pulses set to −12.5 dB and −6.5 dB, respectively. Both the longitudinal ($T_1$) and transverse ($T_2$) nuclear spin relaxation characteristics of our hydrating CPB materials were measured, with the RF pulse sequences employed illustrated in Figure S2. Given the notable content of $Fe_2O_3$ bearing materials (*i.e.* tailings and fly ash) employed in this study, measurements were deliberately perfored at low magnetic field to limit the influence of undesired solid/liquid magnetic susceptibility contrast effects on the $T_2$ relaxation rates of confined water [5]. In general, such effects scale with magenetic field strength and are accentuated in the presence of magnetic impurities [6]. Both $T_1$ and $T_2$ measurements were performed followung to 28 days of hydration. Each measurement was conducted with a constant magnet bore temperature of 30 °C; samples were left within the magnet bore for approximately 15 min prior to each measurement so as to attain thermal equilibrium. We utilised a recycle delay of 10 s and 4 repeat scans across all measurements, which allowed experiments to be completed in 15 mins and 1 min for $T_1$ and $T_2$ measurements respectively, with signal-to-noise ratios > 200. These measurement time scales define the temporal uncertainty of each aquired $\langle T_1 \rangle$ and $\langle T_2 \rangle$ data point in Figure 2 of the main text. It is further assumed that any changes to the CPB surface relaxivity was negligible on the time scale of minutes.

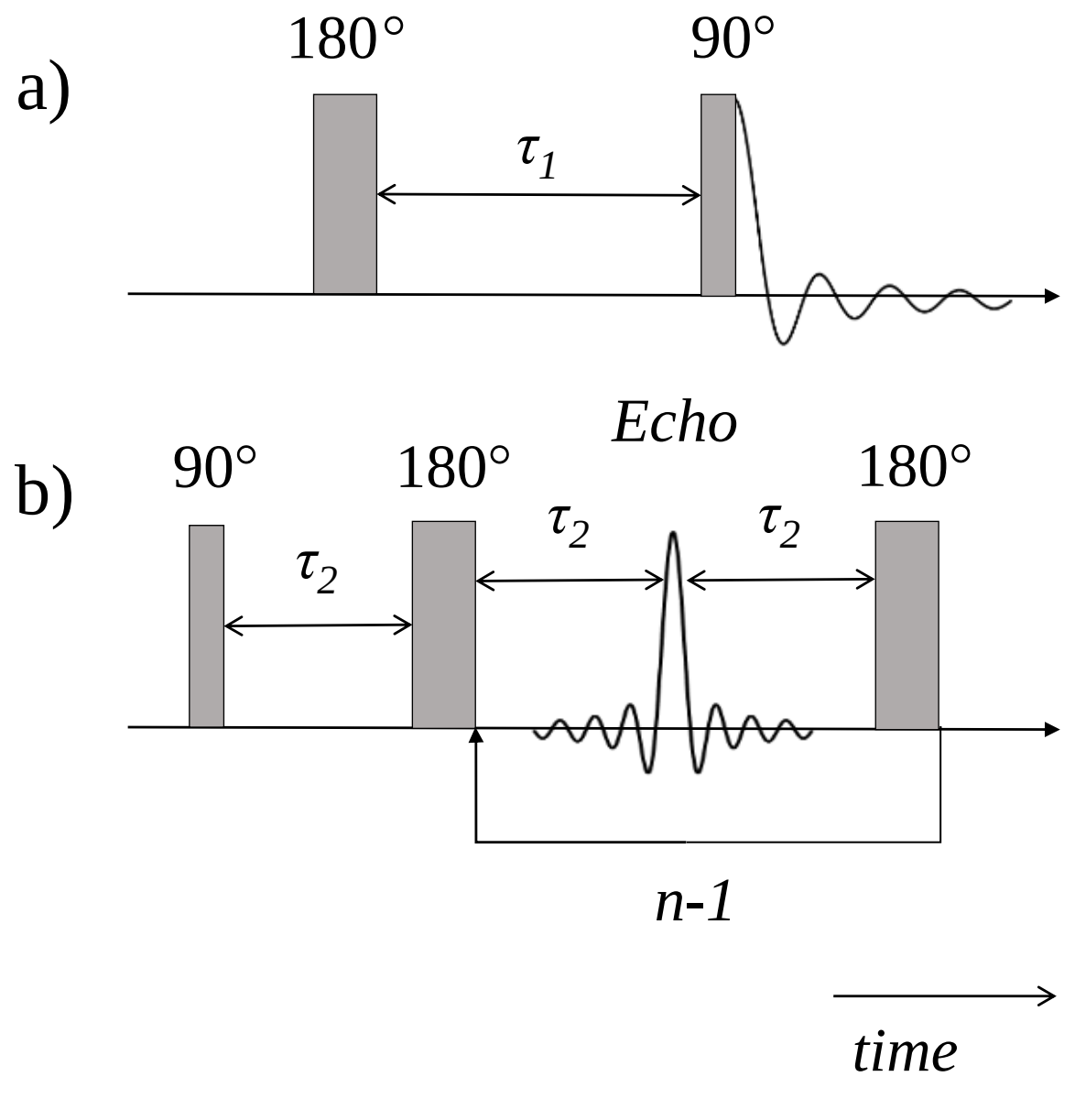

**Figure S2**: Radio frequency (RF) pulse sequence diagrams for a) $T_1$ relaxation time measurements *via* the inversion recovery experiment, and b) $T_2$ relaxation time measurements *via* the CPMG experiment. Vertical bars represent RF pulses; $\tau_1$ represents the variable recovery delays used in the inversion recovery experiment while the CPMG echo time is $t_e = 2\tau_2$.

Longitudinal relaxation times were measured using the standard inversion-recovery (IR) pulse sequence shown in Figure S2a [7]. Starting from thermal equilibrium wherein the sample magnetisation lies parallel to the external magnetic field (conventially the *z*-axis), the IR sequence is initiated by a 180° RF pulse, which rotates the sample magnetisation onto the -*z* axis. Longitudinal relaxation processes are then permitted for a variable recovery period $\tau_1$ during which the sample magnetisation recovers towards thermal equilibrium. Following each recovery period a 90° RF pulse is employed to rotate the sample magnetisation into the transverse plane, allowing the magnitude of the sample magnetisation to be detected [8]. In this work 32 logarithmically-spaced $\tau_1$ recovery times between 0.1 ms and 500 ms were used. The resulting longitudinal relaxation data are described according by a Fredholm integral equation of the first kind [9],

$$\frac{S(\tau_1)}{S(\tau_1 \to \infty)} = \int K_1(\tau_1 T_1) f(T_1)\, \mathrm{dlog}_{10}(T_1) + \varepsilon(\tau_1). \qquad \text{(S1)}$$

Here $S(\tau_1)/S(\tau_1 \to \infty)$ is the normalised NMR signal intensity following each $\tau_1$ delay, the Kernel function $K_1(\tau_1, T_1) = 1 - 2\exp(-\tau_1/T_1)$ describes the expected form of $T_1$ relaxation following the IR experiment [7], and $f(T_1)$ represents the desired distribution of $T_1$ time constants present; $\varepsilon(\tau_1)$ is the experimental noise, assumed Gaussian in shape with zero mean.

Transverse relaxation times were measured using the standard CPMG (Carr-Purcell-Meiboom-Gill) experiment shown in Figure S2b [10, 11]. Starting again from thermal equilibrium the sample magnetisation is excited into the transverse plane *via* a 90° RF pulse, which initiates transverse dephasing of the system's individual nuclear magnetic moments. A series of *n* 180° RF pulses refocusses transverse dephasing of the spin ensamble arising from inhomogeneity across the static magnetic field, generating a train of *n* spin echoes which decay according to $T_2$. In this work we employed *n* = 1000 echoes separated by an echo time of $t_e$ = 100 µs; this short echo time, together with the low static magnetic field strength employed in this study, aimed to limit the influence of undesired transverse relaxation phenomena due to magnetic susceptibility contrast effects at the solid-liquid interface [5]. The resulting relaxation data are again described by a Fredholm integral equation of the first kind [9],

$$\frac{S(nt_e)}{S(0)} = \int K_2(nt_e, T_2) f(T_2)\, \mathrm{dlog}_{10}(T_2) + \varepsilon(nt_e). \qquad \text{(S2)}$$

Here $S(nt_e)/S(0)$ is the normalised NMR signal intensity acquired during each echo, the Kernel function $K_2(nt_e, T_2) = \exp(-nt_e/T_2)$ describes the expected form of $T_2$ relaxation [11] and $\varepsilon(nt_e)$ again represents the experimental noise.

The desired time constant distributions $f(T_1)$ and $f(T_2)$ were obtained *via* a numerical inversion of our acquired experimental data according to Eqs. (S1) and (S2); as this is an ill-posed problem it was necessary to employ Tikhonov regularisation [12] to ensure the stability of the resulting $T_1$ and $T_2$ distributions in the presence of experimental noise, with the smoothing parameter chosen according to the generalised cross-validation method [13]. This robust mathematical inversion technique makes no assumptions about the shape of the resultant probability distributions, and has been successfully implemented across a range of NMR data analysis protocols, including relaxation time distributions in porous media [14, 15] and droplet sizing methods [16]. The inversion algorithm used in this work was written in MATLAB (MathWorks Inc.) and first used by Griffith *et al.* [17].

### Scanning Electron Microscopy

Samples for scanning electron microscopy (SEM) analysis were prepared by mixing only fly ash and OPC (2:1 ratio by mass) to observe key morphological variations throughout the hydration process. We retained the same water content between NMR and SEM measurements by applying the same water-to-binder ratio. Samples were stored for 7 and 14 days accroding to the conditions described above. Samples were then oven dried at 110 °C for 24 hours, crushed and then sieved to below an aperture size of 500 μm. The crushed powders were mounted onto aluminium stubs using adhesive carbon tape and were uniformly coated with a 7 nm thick platinum film to enhance sample conductivity. The same SEM sample preparation procedure was adopted to image the unhydrated fly ash sample shown in Figure 2c of the main text. All imaging utilised a high resolution Zeiss 1555 Field-Emission Scanning Electron Microscope (FESEM) fitted with an X-Max 80 silicon drift X-ray detector (80 mm$^2$) interfaced to Oxford Instruments AztecEnergy software (Oxford Instruments, Oxfordshire, UK). A secondary electron detector operating in high-vacuum mode was employed to observe the morphological characteristics of our samples, with all images acquired using accelerating voltages of 2.5-5 kV, a 9 mm working distance and a 60 μm aperture size.

## 2) Supplementary Results and Discussion

### Data Fitting

All curve fitting was performed using the Curve Fitting Tool in MATLAB (MathWorks Inc.), where non-linear least squared fits were performed using the Trust-Region algorithm. Paramters obtained from the fitting proceedures performed in this work are provided in Tables S2 and S4 below. Table S3 defines the parameters extracted from ref [1] for a hydrating CPB material mixed without using fly ash as a co-binder. These data were obtained from fitting the hydration behaviour of a CPB material exhibiting the same binder content (9 wt%) and water source (tap water) as employed in this work, with the resulting curves shown in purple within Figures 2a and 2b of the main text.

**Table S2**: Parameters obtained by fitting aquired $\langle T_1 \rangle$ and $\langle T_2 \rangle$ data at $t_h < 10$ days to an offset mono-exponential of the form $\langle T_{1,2} \rangle = a_{1,2} \exp\left(-b_{1,2} t_h\right) + c_{1,2}$. The results of this fitting are illustrated in Figures 2a and b of the main text (blue dotted lines) and highlight the rapid decay of our data at short $t_h$. Note that $b_{1,2}$ values were fixed and set to equal those obtained from a full decay-recovery fit across all $t_h$ values; these values can be found in Table S4. Uncertainty ranges correspond with the 95% confidence intervals obtained for each fitted parameter.

| Parameter | Value | Units |
|---|---|---|
| $a_1$ | 6.71 ± 1.23 | ms |
| $c_1$ | 1.48 ± 0.51 | ms |
| $a_2$ | 3.31 ± 0.56 | ms |
| $c_2$ | 1.30 ± 0.24 | ms |

**Table S3**: Parameters obtained from ref [1], obtained by fitting aquired $\langle T_1 \rangle$ and $\langle T_2 \rangle$ data at to an offset mono-exponential of the form $\langle T_{1,2} \rangle = a_{1,2} \exp(-b_{1,2} t_h) + c_{1,2}$. The results of this fitting are illustrated in Figures 2a and b of the main text (purple lines) and highlight the decrased relaxtion decay rates ontained in the absence of fly ash, relative to our current data at short $t_h$. Uncertainty ranges correspond with the 95% confidence intervals obtained for each fitted parameter.

| Parameter | Value | Units |
|---|---|---|
| $a_1$ | 8.26 ± 0.98 | ms |
| $b_1$ | 0.23 ± 0.05 | day$^{-1}$ |
| $c_1$ | 1.67 ± 0.34 | ms |
| $a_2$ | 4.29 ± 0.31 | ms |
| $b_2$ | 0.13 ± 0.02 | day$^{-1}$ |
| $c_2$ | 1.14 ± 0.19 | ms |

**Table S4**: Parameters obtained by fitting aquired $\langle T_1 \rangle$ and $\langle T_2 \rangle$ data to Equation (4) of the main text. Note that we set $b'_{1,2} = 2b_{1,2}$. Uncertainty ranges correspond with the 95% confidence intervals obtained for each fitted parameter.

| Parameter | Value | Units |
|---|---|---|
| $a'_1$ | 15.93 ± 1.74 | ms |
| $a_1$ | 10.90 ± 1.40 | ms |
| $b_1$ | 0.16 ± 0.02 | day$^{-1}$ |
| $c_1$ | 4.37 ± 0.24 | ms |
| $a'_2$ | 7.77 ± 0.85 | ms |
| $a_2$ | 5.47 ± 0.71 | ms |
| $b_2$ | 0.15 ± 0.02 | day$^{-1}$ |
| $c_2$ | 2.83 ± 0.13 | ms |

# Supplementary References